\documentclass{aastex631}  

\UseRawInputEncoding

\shorttitle{Source Imaging of the Multiple Type-II Solar Radio Bursts}
\shortauthors{Vasanth}

\usepackage{underscore}
\usepackage{graphicx}
\usepackage{textcomp}

\graphicspath{{./}{figures/}}
\usepackage{epstopdf} 

\begin{document}

\title{Source Imaging of the Multiple Type-II Solar Radio Bursts and their Multiple CME-Shock Signatures in the Corona}

\correspondingauthor{V. Vasanth}
\email{vasanth.veluchamy@uj.edu.pl}

\author[0000-0002-6056-7899]{V. Vasanth}
\affiliation{Astronomical Observatory of Jagiellonian University, Krakow 30244, Poland \\}






\begin{abstract}
This study reports the meter-wavelength radio imaging of multiple type$-$II bursts occurred on 01 October 2011 observed by LEARMONTH radio spectrograph
in the frequency range of 180 $-$ 25 MHz, with two coronal shocks from two coronal mass ejections (CMEs) from the
nearby active regions lying adjacent to each other. The two CMEs are well observed by STEREO and LASCO FOV.
The eruption is associated with the multi-peaked soft X-ray flare erupted from N16W45 (first and second C3 and C6 class flares) and from N09W04
(third one an M Class flare). The first type$-$II burst starts during the impulsive phase of the first flare
and the second type$-$II burst starts during the peak time of the third flare. The drift rate of the
second type$-$II burst is larger than that of the first type$-$II burst. The EUV images at AIA 211 {\AA} reveal the
shock-like EUV wave structure along the north-western limb around 09:06 UT, the first type$-$II appearing around
the same time (09:06 UT) confirms the existence of the shock wave and as an outward propagating wave along the disk
around 09:55 UT, the second type$-$II appearing around the same time (09:55 UT) confirms the existence of the shock wave.
According to the simultaneous EUV and radio imaging data the radio sources appears along the nose-front of the shock like EUV
wave structure for the first type$-$II burst and for the second type$-$II burst they are located along the shock flank.
The study shows the possibility of two coronal shocks from two CMEs generating the multiple
type$-$II radio bursts. The second event is strongly affected by the projection and scattering effects, the corrections
are achieved by applying spherical cone model and 3D Monte Carlo ray-tracing code for turbulent solar environment at fundamental frequency, the frequency-dependent positional centroid shift for 80 MHz, reached up to $\sim$ 0.12 R$_\odot$ ($\sim$ 120 arcsecs) and for 40 MHz, is about $\sim$ 0.2 R$_\odot$ ($\sim$ 190 arcsecs), including this positional shift demonstrates that the NRH centroids originate from the expanding trailing and lateral flanks of the shock.
\end{abstract}


\section{Introduction}

Solar flares and coronal mass ejections (CMEs)  are the two most efficient energy release processes that play a crucial role in accelerating energetic electrons and triggering shock wave  as well as coronal disturbances. Out of the five types of solar radio bursts, two types (type-II and type-IV radio bursts) are associated with outward propagating coronal structures. Type-II solar radio bursts are narrow band, slow drifting emissive structure observed in the radio dynamic spectrum from higher to lower frequency with a typical drift rate of $\le$ 0.5 MHz/s and with two emission bands, which correspond to the local plasma frequency and its harmonics, sometimes with band-splitting features as well (Payne-Scott et al. 1947; Wild and McCready, 1950; Nelson and Melrose, 1985; Vrsnak et al. 2001; Vrsnak and Cliver, 2008; Prakash et al. 2012; Kong et al. 2012; Feng et al. 2013; Chen et al. 2014; Vasanth et al. 2014; Shanmugaraju et al. 2018; Vasanth, 2024, 2025, 2026). sometimes type-II bursts can also have larger bandwidth (Feng et al. 2015; Zimovets and Sadykov, 2015; Morosan et al. 2022; Vasanth et al. 2025). The type-IV radio bursts are broad band continuum emission with moving and stationary components, for the case of moving type-IV radio bursts the radio sources moves outward from the corona with decreasing frequency over the time, their sources are found to be accompanied by various parts of the ejecta such as plasmoid, advancing front, expanding loop like structures (Boischot, 1958; Riddle, 1970; Sheridan, 1970; Tun \& Vourlidas 2013; Bain et al. 2014; Vasanth et al. 2016, 2019; Carley et al. 2017; Liu et al. 2018; Lv et al. 2021 for latest studies).

Multiple type-II radio bursts are two or more type-II radio bursts that are observed one after the another over a short interval of time, typical within 30 minutes, first reported by Robinson and Sheridan, 1982. Their results imply that a single shock wave intersecting along the boundary of different coronal structures produced the multiple type-II radio bursts. It is important to note that the multiple type-II radio bursts are not observed frequently and are one of the rare phenomena. Several authors have examined the origin, characteristics and properties of the multiple type-II radio bursts (Robinson and Sheridan, 1982; Shanmugaraju et al. 2003, 2005; Subramanian and Ebenezer, 2006; Pohjolainen, Hori, and Sakurai, 2008; Cho et al. 2011; Ramesh, Kathiravan, and Kumari, 2023; Vasanth, 2025, 2026). From the above studies, few authors (Robinson and Sheridan, 1982; Shanmugaraju et al. 2003, 2005; Cho et al. 2011; Vasanth, 2026) supported the single shock wave and their propagating disturbances along the boundaries of coronal structures as the source for generating the multiple type-II radio bursts. While other authors (Subramanian and Ebenezer, 2006; Pohjolainen, Hori, and Sakurai, 2008; Ramesh, Kathiravan, and Kumari, 2023; Vasanth, 2025) comparing the speed of CMEs and speed of multiple type-II radio bursts concluded that the CME fronts and flanks acts as the source for producing the multiple type-II radio bursts.

The chance of occurrence of two kinds of shock waves is explored by few authors using theoretical approach (Karlicky and Odstrcil, 1994), numerical modelling (Magara et al. 2000), simulation studies (Sakai et al. 2005). The possibility of two shock waves generated by solar eruptions, one ahead and a blast wave moving through the CMEs is suggested by Wagner and MacQueen (1983). None of the studies has observed multiple CMEs and their multiple shock waves as cause for the generation of multiple type-II radio bursts. Recently, Vasanth  (2025) reported for the first time, two coronal shock structures from two successive CMEs as the origin for the multiple type-II radio bursts.  The successive type-II bursts were associated with two CMEs originating from the backside limb and thought to be released from the same active region. The study also lacks radio imaging observations.

 The present study reports the multiple type II radio bursts occurred in sequence on 01 October 2011 observed by the LEARMONTH radio spectrograph (Kennewell and Steward, 2003) in the frequency range of 180 - 25 MHz and the radio imaging observations are available from the Nancay Radio Heliograph (NRH: Kerdraon and Delouis, 1997) observations. The EUV and radio imaging observations further allow us to explore the origins of these phenomena. In Section 2, provides the detailed description of the observations at radio and X-ray bands, section 3 provides the description of eruptive structures at EUV, and white-light observations, section 4 furnish the results by comparing the information and features observed in the dynamics spectrum, radio imaging and their relative location on the eruptive structures, section 5 for discussions, and section 6 conclude the results.

\section{Radio Spectral data at Meter wavelength}

The radio dynamic spectrum of the multiple type-II radio bursts recorded on 01 October 2011 observed by the Learmonth radio spectrograph in the frequency range of 180 $-$ 25 MHZ with a time resolution of 3s is shown in Figure 1. The first type-II radio burst starts at 09:06 UT around 170 MHz and drifts to lower frequency and ends at 09:12 UT around 55 MHz in the harmonic band with an average drift rate of 0.3 MHz/s. The second type-II burst starts at 09:55 UT around 160 MHz and ends at 09:57 UT around 90 MHz in the harmonic band with an average drift rate of 0.6 MHz/s, the average drift rate of the multiple type-II bursts are in agreement with the previous studies on type-II bursts (Mann et al. 1996; Vrsnak et al. 2001; Vasanth et al. 2011, 2014), and the previous studies on multiple type-II radio bursts had reported that the first type-II burst drifts faster than the second type-II burst (Robinson and Sheridan, 1982, Shanmugaraju et al. 2005; Cho et al 2011, Vasanth 2025). While in the present study the second type-II burst have faster drift rate compared to the first type-II radio burst. Both the radio bursts appear to have irregular band splitting structure in the harmonic bands and observed with the fundamental and harmonic band structures. It also has a radio continuum emission probably stationary type-IV radio burst after 09:15 UT.

\begin{figure}
\includegraphics[width=0.95\textwidth]{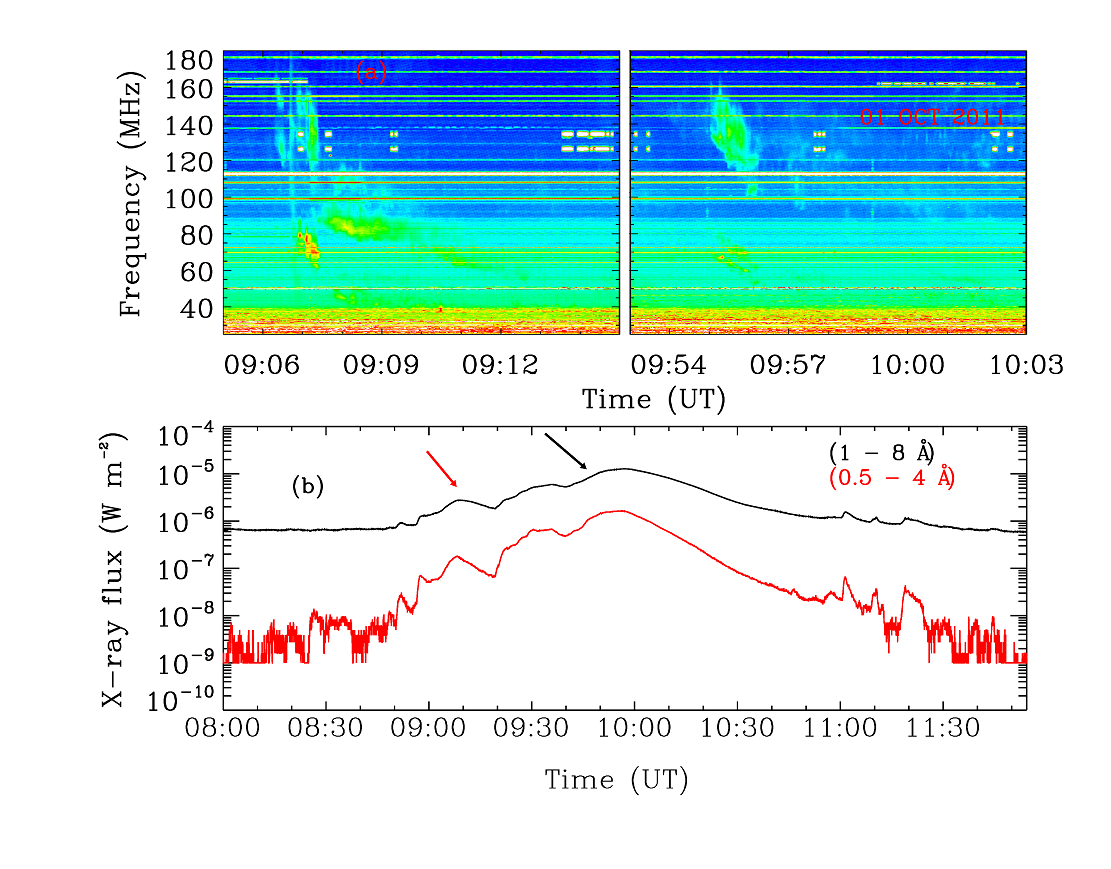}
\break
\caption{(a) The radio dynamic spectrum of the multiple type-II radio bursts recorded by the \textrm{LEARMONTH radio spectrograph} (180 $\--$ 25 MHz) on 01 October 2011 between 09:05 UT $\--$ 09:15 UT $\&$ 09:53 UT $\--$ 10:03 UT, (b) \textrm{GOES} X-ray profile at 0.5 $\--$ 4 and 1 $\--$ 8 {\AA}.
\label{fig:fig1}}
\end{figure}
\hfill \break

The first type II burst is observed in the harmonic band between 09:06 $-$ 09:12 UT, about 6 minutes in the frequency range of 170 $-$ 55 MHz. While the band-splitting is clear until 09:08 UT. Applying 2 X, 3.5 X and 5 X Saito electron density models [Saito et al. 1970], the heights of the type II bursts are estimated by selecting several frequency point on the harmonic band, then they are converted to the fundamental band by dividing the frequency points by 2. The 2 X, 3.5 X and 5 X multipliers on the Saito electron density model are an empirical bracket for coronal structures denser than the quiet Sun such as streamers, active regions and CME-associated density enhancements. For events occurring on the disk, the observations provide only apparent position and projection speed. So density model with different multipliers are used and is compared it with the NRH imaging observations. The estimated heights of the first type-II burst lies in the range of 1.28 $-$ 1.89 $R_\odot$, moving with the average speeds of 1701 $\pm$ 232, 1982 $\pm$ 269 and 2163 $\pm$ 294 km $s^{-1}$ respectively. The second type II burst is observed in the harmonic band between 09:55 $-$ 09:57 UT, about 2 minutes in the frequency range of 160 $-$ 90 MHz. Similarly the estimated heights of the second type-II burst lies in the range of 1.31 $-$ 1.65 $R_\odot$, moving with the average speeds of 1282 $\pm$ 224, 1506 $\pm$ 263 and 1651 $\pm$ 289 km $s^{-1}$ respectively.

The multiple type-II radio bursts were associated with the multi-peaked soft X-ray flux (see Figure 1b), the time corresponding to the first and second type-II bursts are marked by red arrow (first radio burst) and black arrow (second radio burst) in the Figure 1b. The first and second flares are impulsive C3 and C6 class flares located around N16W45 and the C3 flare started at 08:48 UT, peaked at 09:10 UT and ended at 09:20 UT and C6 flare started at 09:20 UT, peaked at 09:38 UT and ended at 09:43 UT. The first type-II burst starts during the impulsive phase of the first flare. The third flare is an M-class flare located around N09W04 started at 9:00 UT, peaked at 09:59 UT and ended at 10:17 UT. The second type-II burst starts during the flare peaking time of the third flare. The multiple solar flares from the different active regions located adjacent to each other are associated with the multiple CMEs . The location of the active regions are shown in Figure 2.
The type$-$II radio sources were imaged by NRH at three different frequency channels (228, 173.2 and 150.9 MHz) with a time resolution of 0.25s, the high resolution data with 0.25s is integrated to 1s for a higher signal$-$noise ratio. The spatial resolution of NRH images depends on the imaging frequency and time, about 1' $-$ 3' at 327 MHz, 3' $-$ 6' at 150.9 MHz.

\begin{figure}
\includegraphics[width=0.95\textwidth]{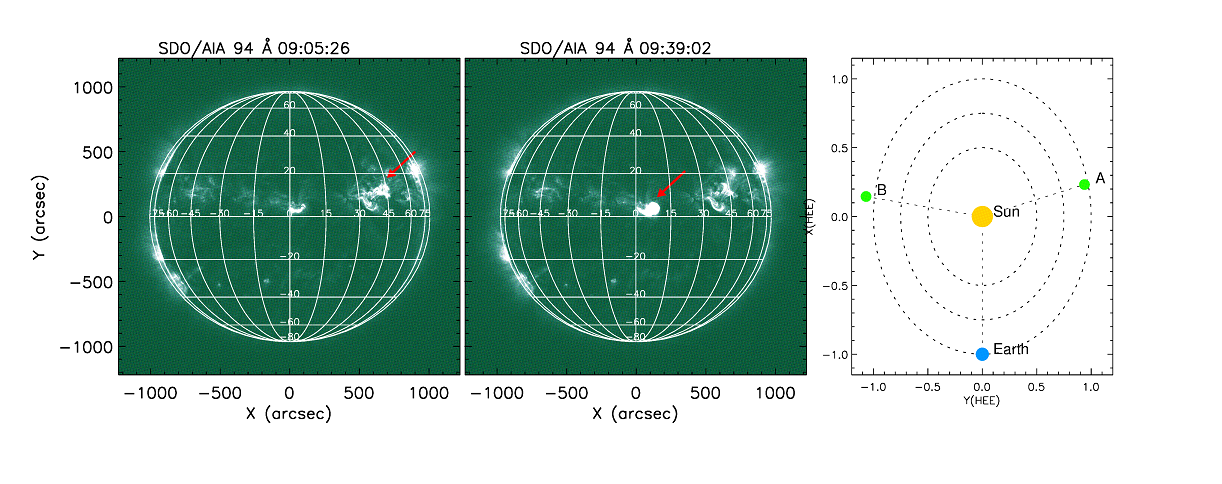}
\caption{The source active region site of the first and second type$-$II radio bursts together with the relative location of twin space craft STEREO and the Earth. The red arrow indicates the active region site.\\
\label{fig:fig2}}
\end{figure}

\section{Eruptions at EUV and White light observations}

The multiple type-II radio bursts were associated with two CMEs originating on the solar disk from different active regions located adjacent to each other according to several instruments. The possible origin and early phase of the CMEs are well observed in the Atmospheric Imaging Assembly (AIA) instruments (Lemen et al. 2012) onboard the Solar dynamics Observatory (SDO) spacecraft (Pesnell et al. 2012) at several EUV wavelengths. The eruptive structures observed at AIA 94, 193 and 211 {\AA} pass-bands are shown using their running difference images in Figure 3. There appears brightening of loop system in the active region AR11302 followed by a violent eruption and the bright structure rapidly moves outward. The bright structure of the eruption is likely associated with the flux rope consisting of the field lines and free magnetic energy to power the eruption. The traces of shock waves are observed in the running difference images at Figure 3 around 09:06 UT for the first radio bursts after it crossing the limb. The second eruption originated from the center of the solar disk in the active region AR11305, there appears brightening of the loop system that grows bigger in size like a bulb and suddenly evolves into a propagating disturbance that form the shock wave around 09:55 UT for the second radio burst. The well-observed shock structure corresponding to first type-II radio burst crossing the limb around 09:07 UT at AIA 193 and 211 {\AA} pass-bands running difference images are shown in top panel of Figure 3. The well observed shock structure on the solar disk corresponding to the second type-II radio burst appearing around 09:55 UT at AIA 193 and 211 {\AA} pass-bands running difference images are shown in bottom panel of Figure 3.

\begin{figure}
\includegraphics[width=0.95\textwidth]{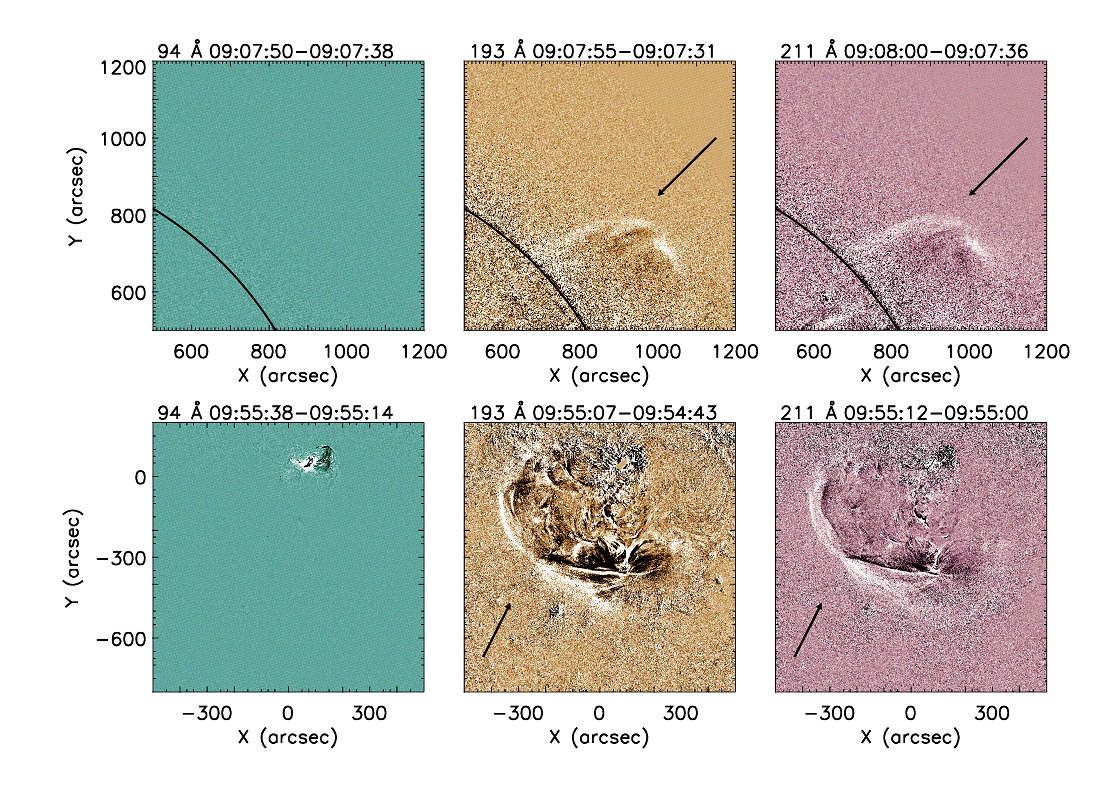}
\caption{The eruptive structures observed by AIA running difference images at 94, 193, and
211 {\AA} for first type$-$II radio burst over the limb indicated with black arrow (top panel), and for the second type$-$II radio burst in the solar disk indicated with black arrow (bottom panel).\\
\label{fig:fig3}}
\end{figure}

\begin{figure}
\includegraphics[width=0.95\textwidth]{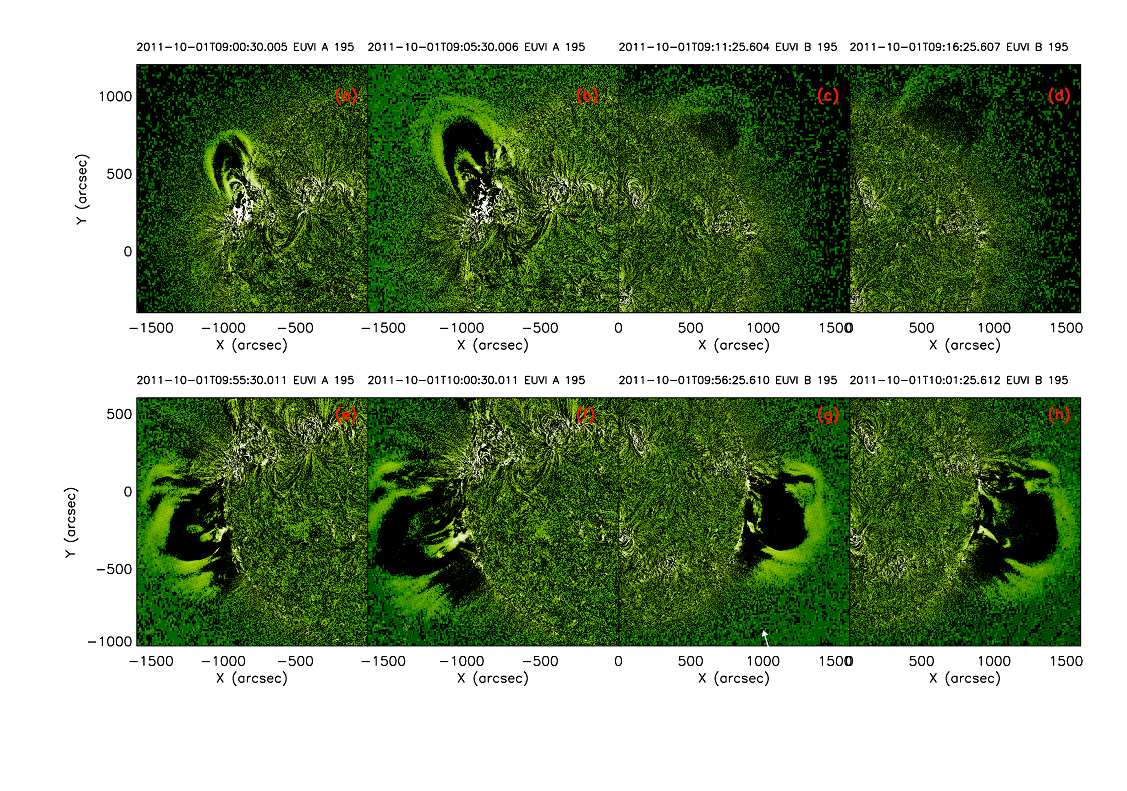}
\caption{Sequence of ST EUVI running difference images at 195 {\AA} showing the eruptive structures for the first type$-$II radio burst (top panel), and the second type$-$II radio burst (bottom panel).\\
\label{fig:fig4}}
\end{figure}

The possible earliest signatures of CMEs are observed by the twin Solar Terrestrial Relations Observatory (STEREO: Kaiser et al 2008) in the Extreme Ultra Violet Imager (EUVI) at 195 {\AA} pass-band  and in the Internally Occulting Refractive Coronagraphs (COR1) FOV. The location of the STEREO instruments are shown in Figure 2. The evolution of CMEs and their shock structure observed in ST-A/B EUVI, ST-A/B COR1 and Large Angle Spectrometric Coronagraph (LASCO) C2 (Brueckner et al. 1995) onboard Solar and Heliospheric Observatory (SOHO) spacecraft (Domingo et al. 1995) SOHO/LASCO observations are shown in Figure 4 and Figure 5. Figure 4 shows the ST-EUVI observations, the first CME is observed in the ST-A EUVI 195 {\AA} images between 09:00 $-$ 09: 10 UT, there appears brightening of loops in the north-eastern limb followed by  a fast propagating waves that moves outward and the EUV structure is observed in the heights between 1.34 $-$ 1.61 R$_\odot$, moving with an average speed of 615 km/s, the heights are measured along the frontal structure of the CME and their earliest shock signature identified by their running difference image around 09:06 UT located at 1.61 R$_\odot$, further confirmed by the presence of type-II burst around 09:06 UT. It is detected in the ST-B EUVI between 09:06 $-$ 09:21 UT as a diffuse structure that originated on the backside of the solar limb (See top panel of Figure 4). The EUV structures are observed in ST-B in the height range of 1.19 $-$ 1.67 R$_\odot$, moving with an average speed of 546 km/s. The first CME appears in the ST-A COR1 FOV between 09:15 $-$ 09:55 UT in the height range of 1.75 $-$ 3.66 R$_\odot$, moving with an average speed of 556 km/s and it appears in the ST-B COR1 FOV between 09:21 $-$ 09:56 UT in the height range of 1.74 $-$ 3.021 R$_\odot$, moving with an average speed of 411 km/s.

The second CME is observed in the ST-A EUVI 195 {\AA} images from the backside of the solar disk around 09:40 $-$ 09:55 UT, there appears sudden brightening and freely propagating wave in the heights between 1.18 $-$ 1.68 R$_\odot$, moving with an average speed of 384 km/s. It is observed in the ST-B EUVI between 09:41 $-$ 10:01 UT partially on the north-western limb and partially backside (see bottom panel of Figure 4). The EUV structures are observed in the height range of 1.16 $-$ 1.68 R$_\odot$,moving an average speed of 322 km/s. The second CME appears in the ST-A COR1 FOV between 10:00 $-$ 10:50 UT in the height range of 1.75 $-$ 4.34 R$_\odot$, moving with an average speed of 544 km/s and in the ST-B COR1 FOV between 10:01 $-$ 10:51 UT in the height range of 1.73 $-$ 4.08 R$_\odot$, moving with an average speed of 608 km/s (see bottom panel of Figure 5). The CME appears in LASCO C2 FOV around 09:18 $-$ 11:24 UT (first CME) and between 10:36 $-$ 12:24 UT (second CME).

\begin{figure}
\includegraphics[width=0.95\textwidth]{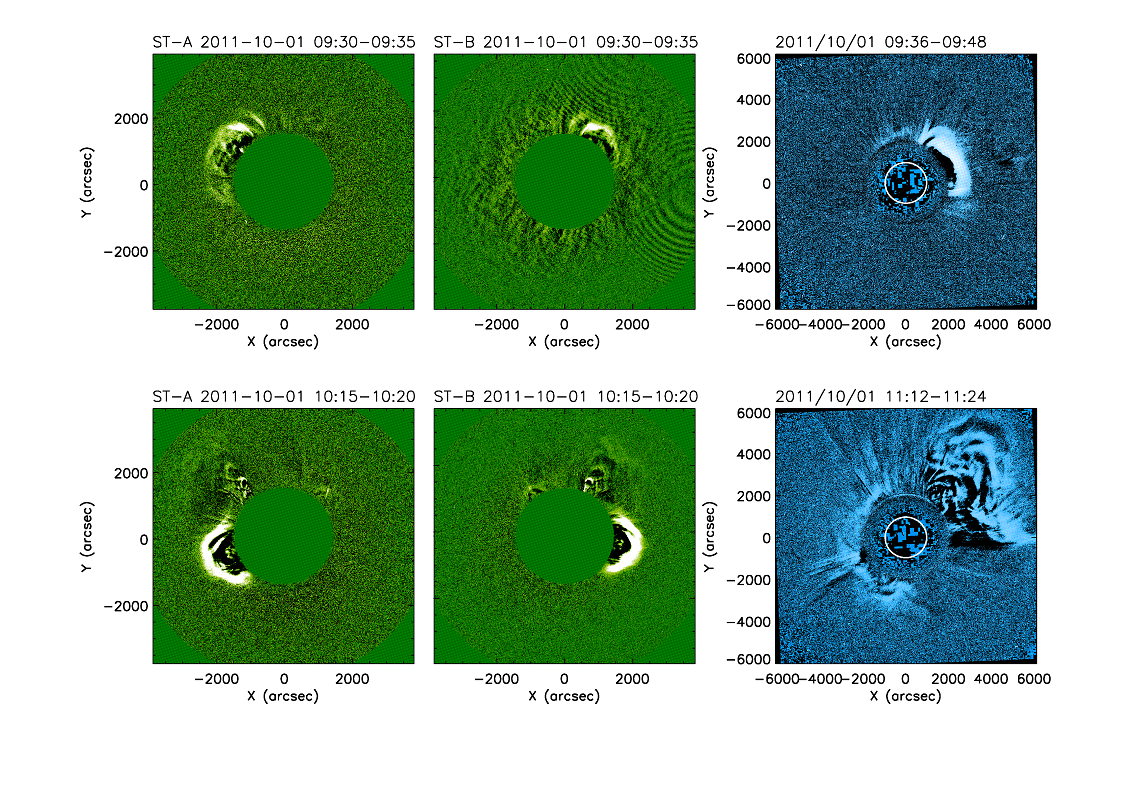}
\caption{The ST COR1 and LASCO-C2 running difference images showing the eruptive CMEs for first type$-$II radio burst (top panel) and the second type$-$II radio burst (bottom panel).\\
\label{fig:fig5}}
\end{figure}

\section{Radio Imagery and their correlation to the eruptive structures}
The highly sensitive NRH radio imaging observations are used to identify the temporal and spatial position of the radio sources that move outward from the eruptive site. We can also utilize the NRH observations to locate the onset of radio emissions at higher frequencies that are not recorded in the radio dynamic spectrum The high-frequency onset is not recorded by LEARMONTH because it lies 48 MHz above the instrument's 180 MHz upper frequency limit, Further high-cadence NRH imaging helps to identify the existence of radio burst at higher frequencies and fills the gap. The NRH radio imaging data can be used to identify and differentiate the different kinds of radio bursts by their shifts of the total flux maximum to subsequent frequencies with time. The NRH provides the radio imaging at 9 different frequencies between 450 $-$ 150 MHz, here we used the last three frequency channels (228, 173.2 and 150.9 MHz). The present study uses the NRH imaging data with a temporal resolution of 0.25 s and integrate them to 1s for better signal to noise ratio. The evolution of the dominant sources is compared with the features observed in radio dynamic spectrum.

The temporal evolution of radio sources at different NRH frequencies 228, 173.2 and 150.9 MHz are shown in Figure 6. The contour represents 80\%, 90\% and 95\% of the maximum brightness temperature (T$_{Bmax}$). The radio sources are short-lived at highest observing frequency and long-lived at lowest observing frequency. This pattern of outward motion of radio source from highest frequency channel to lowest frequency channel with time indicates that we are observing the drifting type-II radio burst. At later time, the observation of radio source stationed at a fixed point on the eruptive site indicates it corresponds to the background continuum emission (stationary type-IV radio burst).

\begin{figure}
\includegraphics[width=0.95\textwidth]{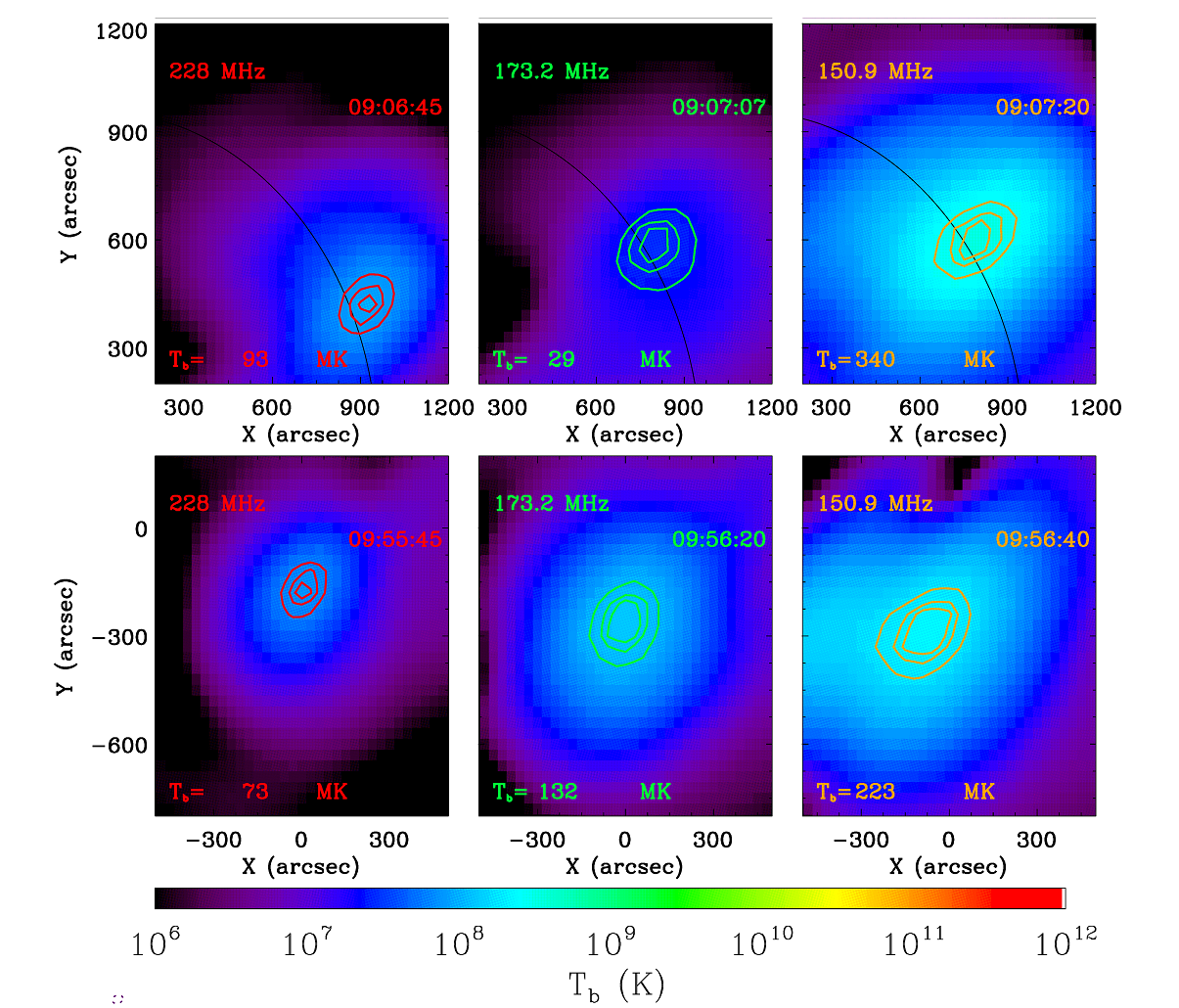}
\break
\caption{The temporal and spatial evolution of Type$-$II radio sources at different NRH frequencies. Top panel showing the radio sources for the first type$-$II radio burst and the bottom panel showing the radio sources for the second type$-$II radio burst. The radio sources are represented by 80\%, 90\% and 95\% T$_{Bmax}$ contours.\\
\label{fig:fig6}}
\end{figure}

The first type-II burst appeared in the NRH FOV around 09:06:22 UT at 228 MHz. Later it appeared at successive lower frequencies. At 173.2 MHz, the T$_{Bmax}$ is 6 x 10$^8$ K at 09:06:44 UT, whereas it is 2 x 10$^8$ K at 09:06:43 UT at 228 MHz and is 6 x 10$^9$ K at 09:07:06 UT at 150.9 MHz. Similarly, the second type-II burst appears in the NRH FOV at 09:55:03 UT at 228 MHz. Later on it appeared at successive lower frequencies. At 173.2 MHz, the T$_{Bmax}$ is 1 x 10$^8$ K at 09:55:07 UT, whereas it is 7 x 10$^7$ K at 09:55:04 UT at 228 MHz and is 1 x 10$^{10}$ K at 09:55:29 UT at 150.9 MHz. The T$_{Bmax}$ reaches higher values at lower frequencies. The type-II radio sources can be identified from 228 MHz to lower frequency channels one after the other with some delay consistent with the drifting motion from higher to lower frequencies observed in the radio dynamic spectrum.The radio sources disappearing at high altitude indicate that the type-II emission had either ceased or moved out of the NRH FOV.

In Figure 7, the NRH radio sources from 228 - 150.9 MHz were over plotted over the temporally closest SDO/AIA 211 {\AA} running difference images to examine the temporal and spatial evolution of radio emissions relative to the eruptive structures. In the case of type-II radio burst, the observed NRH data shows the shift of radio sources from higher to lower frequencies with time. The radio sources corresponding to the first radio bursts are located at the nose-front of the EUV-like shock wave structure, the radio sources at different frequencies do not overlap with each other, they are spatially dispersed and at later stage the radio sources move towards the shock flank, and for the second type-II radio burst the radio sources at different frequencies are spatially dispersed and are located at the shock flank. The first type-II burst radio sources appears at the nose-front of the EUV-like shock wave and for the second type-II burst the radio sources are located at the flank.

The active region appears on the disk, and the observations provides only the apparent position and projection speed. To determine the true dynamic heights of the Type II sources and correct for line-of-sight projection effects, we applied a standard heliocentric deprojection algorithm to the high-cadence NRH observations. It adopts a spherical cone model, treating the emission centroids as localized features that propagate radially outward along a linear trajectory relative to the EUV site/structures. The estimated projection and the approximated displacement of radio sources for the first burst at 228 MHz is about 46 - 135 arcsecs and for the 173.2 and 150.9 MHz it is about 75 - 200 arcsecs. Further the radio sources align well with 2 X Satio electron density model. For the second radio radio burst, 228 MHz radio sources are projected and displaced by about 30 - 160 arcsecs, 173.2 MHz by about 20 - 100 arcsecs and 150.9 MHz by about 3-13 arcsecs respectively. The second burst aligns well with base density model without multiplier. To account for turbulent solar environment, we subsequently utilized the 3D Monte Carlo ray-tracing code (ray_new.pro; Kontar et al. 2019) to simulate anisotropic scattering and refraction at the corresponding plasma frequency layers.

The simulation with a standard background density profile (N=1) and realistic shock-induced turbulence ($\epsilon$=0.12, $\alpha$=0.20) at fundamental frequency, for the second burst the frequency-dependent positional centroid shift for 80 MHz, reached up to $\sim$ 0.12 R$_\odot$ ($\sim$ 120 arcsecs) and for 40 MHz, it is about $\sim$ 0.2 R$_\odot$ ($\sim$ 190 arcsecs). Incorporating these scattering shift vectors into our final spatial grid definitively demonstrates that the NRH centroids originate from the expanding trailing and lateral flanks of a 3D shock bubble. Then, for the first radio burst the estimated frequency-dependent positional centroid shift for 80 MHz, is $\sim$ 0.06 R$_\odot$ ($\sim$ 50 arcsecs) and for 40 MHz, it is about $\sim$ 0.09 R$_\odot$ ($\sim$ 90 arcsecs). The first radio burst is far less affected and the second events is strongly affected by projection and scattering effects.

\begin{figure}
\includegraphics[width=0.95\textwidth]{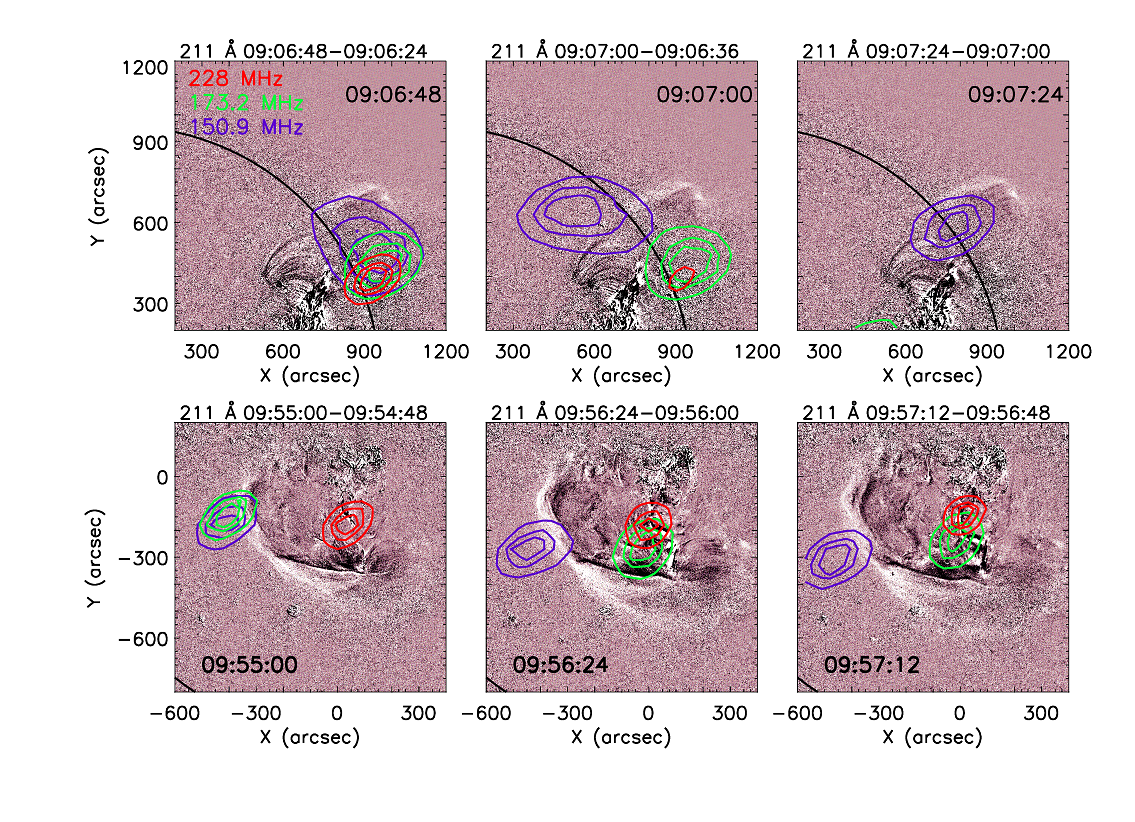}
\caption{The temporal evolution of NRH radio sources superposed onto the closest-in-time AIA 211 {\AA} running difference images for the first type$-$II radio burst (top panel) and the second type$-$II radio burst (bottom panel). The radio sources are represented by 80\%, 90\% and 95\% T$_{Bmax}$ contours.\\
\label{fig:fig7}}
\end{figure}

\section{Discussion}
 Using multi-wavelength observations this study presents robust evidence supporting the existence of two coronal shock waves from two CMEs generating the multiple type-II radio bursts. The two CMEs originated from the nearby active regions lying adjacent to each other and are associated with multiple flares, a C3 class flare from N16W45 and a M1 class flare from N09W04. The first type-II radio burst started during the impulsive phase of the first flare and the second type-II radio burst during the flare peaking time of the second flare. The two shocks were observed by AIA 211 {\AA} running difference images and ST-EUVI images around 09:06 UT and 09:55 UT (see Figure 2), further confirmed by the existence of two type-II bursts during the same time. The availability of NRH radio imaging observations allows us to explore the possible source location on the eruptions, the radio contours at several NRH frequencies superposed over the closest AIA 211 {\AA} running difference images show that the first type-II radio burst is produced along the CME shock-front and the second type-II radio burst along the CME shock-flank. The temporal and spatial occurrence of two CMEs aligning with the existence of two type-II radio bursts favours piston-driven shocks as the possible driving mechanism for generating the multiple type-II radio bursts.

 Multiple type-II radio bursts are two or more type-II burst components emerging in order, one after another over a short interval of time with distinct drift rates, the distinct drift rates suggesting an inhomogeneous plasmas, and the energetic electrons are accelerated at different coronal structures with different density gradients and distinct propagating speeds from two separate shocks and they also might be arising from different source regions (Shanmugaraju et al. 2003, 2005, Subramanian and Ebenezer, 2006, Reiner et al. 2001). Both flares and CMEs can produce shock waves, the presence of type-II radio burst during a flare and or CMEs process confirms the existence of shock waves (Vrsnak and Cliver, 2008).

 The observation of multiple type-II radio bursts suggests the propagation of multiple shock waves in the corona. Robinson and Stewart (1985) suggest that the eruptive events with several flares are more likely to produce multiple type-II radio bursts. But most of the previous studies on multiple type-II radio bursts lack the observational evidence of two coronal shocks and their observational results support single shock waves and their propagating disturbances interacting with the different regions of the coronal structures as the cause for generating the multiple type-II radio bursts (Robinson and Sheridan, 1982, Shanmugaraju et al. 2003, 2005, Subramanian and Ebenezer 2006, Cho et al. 2011).  Therefore, most of the previous studies supported single shock wave with their radio emitting region located on the CME-shock front or CME-shock flank or along the streamer interacting regions.

 Recently, Vasanth (2025) for the first time reported the direct observational evidence on the existence of two coronal shocks from two CMEs generating the multiple type-II radio bursts. The two CMEs are ejected from the same active region. But still they lack radio imaging observations, so they cannot provide the precise radio source location. The present study has the radio imaging observations from NRH and provides the precise source locations over the eruptive structures. The occurrence of two distinct coronal shock waves from two distinct drivers mainly depends on the properties of local ambient medium, as well as their characteristics in the corona. More observational and detailed case studies on multiple type-II radio bursts are required to further explore the potential of multiple shocks scenario with multiple CME eruptions to infer the coronal conditions triggering it.

 \section{Conclusions}
    This study demonstrates the possibility of observing multiple shocks from multiple CMEs generating the multiple type-II radio bursts observed on 01 October 2011 by LEARMONTH radio spectrograph in the frequency range between 180 $-$ 25 MHz. The type-II bursts are observed in both fundamental and harmonic band and the present study used the harmonic band for the analysis. The eruption is associated with the multi-peaked soft x-ray flare energy release processes from N16W45 the C3 and C6 class flares and N09W04, an M1 class flare lying adjacent to each other. The first type-II burst is observed from 170 $-$ 55 MHz between 09:06 $-$ 09:12 UT, with an average drift rate of 0.3 MHz/s. The second type-II radio burst is observed from 160 $-$ 90 MHz between 09:55 $-$ 09:57 UT, with an average drift rate of 0.6 MHz/s. The two type-II bursts have different drift rates and the drift rate of second type-II burst is larger than the first type-II radio burst. The different drift rates indicate different source regions$/$locations.

    The radio imaging observations from NRH allow us to explore the possible source regions$/$locations. According to the simultaneous EUV and radio imaging data the radio sources initially appears along the nose-front of the shock-like EUV wave structure for the first type-II radio burst and later shifts to the shock flank and, for the second type-II radio burst they are located along the shock flank. Therefore the two coronal shocks from two CMEs generated the multiple type-II radio bursts.

    The radio bursts originated from the eruptive structures located on the disk and therefore, the observations provides only the apparent position and projection speed. Therefore, to correct for projection effects, we applied a standard heliocentric deprojection algorithm to NRH observations, adopting a spherical cone model. The estimated projection of radio sources for the first burst lies in the range of 45 - 200 arcsecs and agrees well with the 2 X Saito electron density model. For the second radio radio burst, the radio sources projection lies in the range of 3 - 160 arcsecs and agrees well with base density model without multiplier. Further, the anisotropic scattering and refraction of radio sources at turbulent solar environment are simulated for standard background density profile (N=1) for fundamental frequency and realistic shock-induced turbulence ($\epsilon$=0.12, $\alpha$=0.20), using 3D Monte Carlo ray-tracing code (ray_new.pro; Kontar et al. 2019) at the corresponding plasma frequency layers. For the second burst, the positional centroid shift for 80 MHz, is $\sim$ 0.12 R$_\odot$ ($\sim$ 120 arcsecs) and for 40 MHz, it is about $\sim$ 0.2 R$_\odot$ ($\sim$ 190 arcsecs). These scattering shift demonstrates that the NRH centroids originate from the expanding trailing and lateral flanks of a 3D shock bubble. For the first radio burst the estimated frequency-dependent positional centroid shift for 80 MHz, is $\sim$ 0.06 R$_\odot$ ($\sim$ 50 arcsecs) and for 40 MHz, it is about $\sim$ 0.09 R$_\odot$ ($\sim$ 90 arcsecs). The first radio burst is far less affected and the second events is strongly affected by projection and scattering effects.

\begin{acknowledgments}
This work was supported by POB Anthropocene research program of Jagiellonian University, Krakow, Poland.  I would like to thank Dr.Yao Chen for the fruitful discussions. We thank the LEARMONTH radio spectrograph, NASA, NOAA, NRH, STEREO/SECCHI, and SOHO/LASCO teams for providing the data and LASCO CME catalog generated and maintained by the Center for Solar Physics and Space Weather, the Catholic University of America, in cooperation with the Naval Research Laboratory and NASA.
\end{acknowledgments}

\bibliographystyle{aasjournal}



\end{document}